 \documentclass[final,5p,times,twocolumn]{elsarticle}

\usepackage{amsmath,amssymb,amsfonts}
\usepackage{mathrsfs}
\usepackage{chemformula}
\usepackage{color}
\usepackage[normalem]{ulem}
\usepackage{stmaryrd}
\usepackage{diagbox}

 \biboptions{sort&compress}

\newcommand{\beq}{\begin{equation}}
\newcommand{\eeq}{\end{equation}}

\newcommand{\ep}{\varepsilon}
\newcommand{\Pra}{\textit{Pr}}
\newcommand{\mb}[1]{\mathbf{#1}}
\newcommand{\rr}{\mathfrak{r}}
\newcommand{\mm}{\mathfrak{m}}

\newcommand{\mc}[1]{\mathcal{#1}}

\journal{Proceedings of the Combustion Institute}

\begin{document}

\begin{frontmatter}

\title{Flame dynamics and Markstein numbers in Hele-Shaw cells and porous media   under Darcy's law}

\author{Prabakaran Rajamanickam and Joel Daou}
\address{Department of Mathematics, University of Manchester, Manchester M13 9PL, United Kingdom}

\begin{abstract}
The propagation of premixed flames in narrow Hele-Shaw cells and permeable porous media is governed by Darcy's law, leading to hydrodynamic behaviour distinct from conventional flames. This study investigates the role of confinement on flame dynamics, focusing on the associated Markstein numbers  with arbitrary Lewis numbers. A hydrodynamic model treating the flame as a discontinuity surface is presented, in which the burning rate depends on curvature and tangential flow strain, characterised by two Markstein numbers $\mc M_c$ and $\mc M_t$. A major finding is that $\mc M_c \neq \mc M_t$ under Darcy's law, as the law permits tangential velocity discontinuities  at the flame front due to viscosity variations. Additionally, a third Markstein number $\mc M_g$ associated with gravity also emerges uniquely under Darcy's law. The Darcy-specific effects vanish in purely radial flows but are important for strained flames. In planar counterflows, for instance, the strain rate jump across the flame is dictated by the unburnt-to-burnt viscosity ratio $\mm$ rather than the density ratio $\rr$, a dramatic departure from conventional behaviour. The influence of confinement on the combined hydrodynamic instabilities of planar flames, namely Darrieus--Landau, Saffman--Taylor, and Rayleigh--Taylor instabilities, is discussed. Weakly nonlinear dynamics under strong confinement is found to follow a Michelson--Sivashinsky equation with modified coefficients (long-wave instability), while under moderate confinement, Ginzburg--Landau dynamics (finite-wavenumber instability) is found to apply.  Strong confinement amplifies the Darrieus--Landau instability, enhancing hydrodynamic coupling in conjunction with augmented streamline refraction caused by tangential velocity discontinuities.
\end{abstract}


\end{frontmatter}

\section{Introduction\label{sec:introduction}} 

Premixed flames in permeable porous media differ fundamentally from conventional flames,  because the hydrodynamics in these systems are governed by Darcy's law. Similarly, flames propagating in the narrow gap of a Hele-Shaw cell with nearly adiabatic walls may also be qualitatively modelled using Darcy's law, as has been done in several recent studies~\cite{fernandez2018analysis,martinez2019role,rajamanickam2024effect,daou2025hydrodynamic,chen2025three}. Darcy's law accounts for frictional momentum loss, and a rational justification for its application is provided in~\cite{daou2025hydrodynamic}. Consequently, physical parameters like the permeability of the porous medium or the width of the Hele-Shaw channel become critical in dictating the flame's behaviour.

The pioneering work of Joulin and Sivashinsky~\cite{joulin1994influence}  laid the foundation for understanding confinement effects by developing a stability analysis for planar flames based on an Euler--Darcy model. More recent studies have extended this framework, incorporating the effects of imposed flow~\cite{miroshnichenko2020hydrodynamic,daou2025hydrodynamic}. Specifically, the study of~\cite{daou2025hydrodynamic} derived a dispersion relation for the stability of planar flames under Darcy's law, involving two Markstein numbers. The Markstein numbers, which characterise the burning rate of curved and stretched flames, are treated as prescribed parameters in the dispersion relation. Explicit formulas for Markstein numbers in terms of physico-thermal properties were first derived from the Navier--Stokes equations with one-step chemistry by Clavin and Williams using large activation-energy asymptotics~\cite{clavin1982effects}. These results were subsequently extended and generalised in various contexts~\cite{clavin1982effects,pelce1988influence,matalon1982flames,clavin1983premixed,clavin1983influence,clavin1985effect,keller1994transient,matalon2009multi,bechtold2024hydrodynamic}. Alternatively, Markstein numbers may be extracted from known dispersion relations obtained through   numerical simulations~\cite{daou2025hydrodynamic} or   experimental measurements~\cite{al2019darrieus}, or from flame propagation data obtained numerically and experimentally in simple configurations~\cite{chen2011extraction}. A key open question is whether Markstein numbers, and the underlying relations linking them to flame stretch, curvature, and propagation, remain applicable under strong confinement, or if they require fundamental revision. Despite this uncertainty, and in the absence of more rational alternatives, studies in the literature~\cite{al2019darrieus,radisson2022forcing} often adopt Markstein-number expressions derived for unconfined flames even in strongly confined settings.

The primary objective of this work is to clarify how confinement, through Darcy's law, alters the Markstein numbers and the associated relations between flame curvature, flow strain, and flame propagation, and consequently influences the resulting dynamics of premixed flame fronts. The present work extends our recent study~\cite{rajamanickam2024hydrodynamic}, which was restricted to equidiffusive flames ($Le=1$), to the more general case of arbitrary Lewis numbers. This generalisation is essential for capturing the coupling between differential diffusion and the   hydrodynamic features associated with   Darcy flows considered herein. To this end, we analyse several canonical problems, including radially symmetric flames, planar counterflow flames, and the hydrodynamic flame instabilities, to highlight   distinctive phenomena that emerge under confinement.

\section{Hydrodynamic model of premixed flames and Markstein numbers under Darcy's law}

Throughout the paper, we shall use the laminar flame thickness $\delta_L=D_{T,u}/S_L$ as the characteristic length scale and the flame residence time $\delta_L/S_L$ as the characteristic time scale; $D_{T,u}$ is the thermal diffusivity in the unburnt gas and  $S_L$ here need not correspond to the standard laminar flame speed but represents an effective planar flame speed in porous media or a speed that accounts for flame curvature in the wall-normal direction in Hele-Shaw channels, as discussed in~\cite{joulin1994influence,daou2025hydrodynamic,chen2025three,dejoan2023flame}. This 'effective' flame speed approach is also conceptually similar to the classical formulations used for multiphase and dusty-gas combustion~\cite{seshadri1992structure,demir2017towards}. For simplicity, we neglect heat loss effects and assume that the reacting mixture is fuel lean having a constant Lewis number $Le$.

We introduce the non-dimensional density $\rho(\theta)$, the product of density and thermal diffusivity $\lambda(\theta)$ and the viscosity $\mu(\theta)$ (or viscosity-to-permeability ratio in porous media), all scaled by unburnt-gas values.  The non-dimensional temperature is denoted by $\theta \in [0,1]$. A commonly adopted model assumes $\rho=1/(1+q\theta)$ and $\mu=\lambda=(1+q\theta)^{0.7}$ where $q$ quantifies the heat release~\cite{daou2025hydrodynamic}. The resulting ratio of unburnt-to-burnt gas density, $\rr=\rho_u/\rho_b$, and the  ratio of unburnt-to-burnt viscosity/permeability, $\mm=(\mu_u/\kappa_u)/(\mu_b/\kappa_b)$, are then given by
\begin{equation}
    \rr = 1 + q, \qquad \mm =(1+q)^{-0.7}. 
\end{equation}
Here, $\kappa$ is the   permeability with values $\kappa_u$ and $\kappa_b$ in the unburnt and burnt gas regions, respectively. If the permeability is constant, then   $\mm=\mu_u/\mu_b$ is just the ratio of unburnt-to-burnt viscosity. This is the case, for example,  in a Hele-Shaw channel where $\kappa = h^2/12$,  with $h$ denoting the channel width. In this work, the   hydrodynamics is described by
 Darcy's law rather than by the Navier--Stokes equation.  In our non-dimensional notation, Darcy's law takes the form
\begin{equation} \label{eq:FullDarcy}
\mu \mb v = -\nabla p + \rho \mb g ,   
\end{equation} 
where $p$ is the pressure scaled by $\delta_L S_L\mu_u/\kappa_u$  and $|\mb g|=\rho_u g \kappa_u/\mu_u S_L$ is a non-dimensional measure of the gravitational force strength.

\begin{figure}[h!]
\centering
\includegraphics[width=0.28\textwidth]{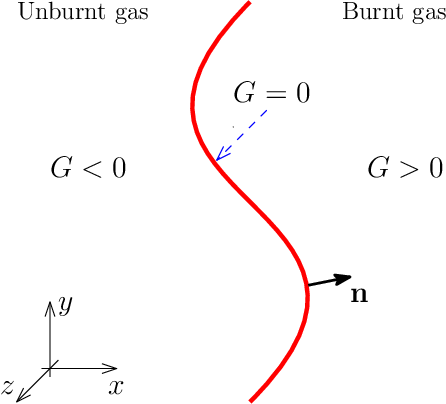} 
\vspace{10 pt}
\caption{Schematic   of a curved premixed flame. The flame front, defined by the level set $G(\mb x,t)=0$, is assumed to have a wrinkling length scale much larger than its thickness.} 
\label{fig:sch}
\end{figure}

The premixed flame is modelled as a discontinuity surface, following the approach of Navier--Stokes-based studies such as \cite{clavin1982effects, pelce1988influence, matalon1982flames}. The flame front may be represented as the level set $G(\mb x,t)=0$ with $G<0$ corresponding to unburnt gas and $G>0$ to burnt gas, as illustrated schematically in Fig.~\ref{fig:sch}.  On either side of the flame, the flow is incompressible and satisfies Darcy's law~\eqref{eq:FullDarcy} such that 
\begin{align}
    &\nabla\cdot\mathbf{v} = 0, \quad \mathbf{v} = -\nabla p + \mathbf{g} \quad (G < 0), \label{darcyunburnt}\\[4pt]
    &\nabla\cdot\mathbf{v} = 0, \quad \frac{\mathbf{v}}{\mm} = -\nabla p + \frac{\mathbf{g}}{\rr}     \quad (G > 0) . \label{darcyburnt}
\end{align}
 As a consequence of these two equations, the pressure satisfies Laplace's equation $\nabla^2 p = 0$ for $G \neq 0$.  At the flame front located at $G = 0$, the conservation of mass and pressure continuity imposed by the jump conditions
\begin{equation}
    \llbracket \rho(\mathbf{v}\cdot\mathbf{n} - U_f) \rrbracket = 0, 
    \qquad 
    \llbracket p \rrbracket = 0 
    \label{interfacedarcy}
\end{equation}
hold to a first approximation. Here, $\llbracket \varphi \rrbracket \equiv \varphi|_{G=0^+} - \varphi|_{G=0^-}$, 
$\mathbf{n} = \nabla G / |\nabla G|$ is the unit normal directed toward the burnt gas, 
and $U_f = -(\partial G / \partial t)/|\nabla G|$ denotes the normal flame speed in the laboratory frame. We note that the pressure continuity follows readily from integrating Eq.~\eqref{eq:FullDarcy}, which is valid throughout the domain,  across the flame front. The model is completed by the kinematic condition (the $G$-equation) applied on the unburnt side $G = 0^-$, namely
\begin{equation}
    \frac{\partial G}{\partial t} + \mathbf{v} \cdot \nabla G = \dot m |\nabla G|. \label{kinematic}
\end{equation}
In this equation, $\dot m$ denotes the burning rate evaluated on the unburnt-gas side of the flame (scaled by $\rho_u S_L$). Its determination, together with the Markstein numbers it involves, is an important outcome of the present investigation.  To this end,  we have carried out   a multi-scale asymptotic analysis using a one-step Arrhenius chemistry model with large activation energy, for flames whose thickness is small compared to the hydrodynamic/geometric scale. The analysis generalises that of~\cite{daou2025hydrodynamic} to non-unit Lewis numbers. Although the procedure closely follows~\cite{daou2025hydrodynamic}, it is rather lengthy. Therefore, we record here only the main findings, in the form of explicit formulas for the burning rate $\dot m$ and the associated Markstein numbers, which are defined here more carefully than in~\cite{daou2025hydrodynamic}. We also apply these results to canonical problems, both to elucidate the effect of confinement and to illustrate the use of the theory for combustion applications. 

We begin by presenting the results for the case \(\mathbf{g} = \mathbf{0}\), before turning to the more general case with gravity. Specifically, for \(\mathbf{g} = \mathbf{0}\), the resulting formula for the burning rate can be written as
\begin{align}
     \dot m =
     1- \mc M_c \mathbb K + (\mc M_c - \mc M_t)\nabla_t\cdot\mb v_t   \label{burningrate2}  
\end{align}
or, equivalently, 
\begin{align}
     \dot m  =  1+ \mc M_c (1 - \mb v\cdot \mb n)\nabla\cdot \mb n - \mc M_t \nabla_t\cdot \mb v_t \label{burningrate}    \,.
\end{align}
Here $\mathbb K=-\nabla\cdot \mb n -\mb n\mb n:\nabla \mb v$ denotes the usual kinematic flame stretch~\cite{matalon2018darrieus,clavin2016combustion}, and
$\nabla_t\cdot \mb v_t= - \mb n\mb n:\nabla\mb v - (\mb v\cdot \mb n)\nabla\cdot \mb n$ is the surface divergence of the tangential velocity $\mb v_t=(\mb I-\mb n\mb n)\mb v$, where the above expression for $\nabla_t\cdot \mb v_t$ follows after using $\nabla\cdot \mb v=0$.  Furthermore, $\mc M_c$ and $\mc M_t$ as defined in formulas~\eqref{burningrate2} and~\eqref{burningrate} are referred to as
the \textit{curvature Markstein number} and the  \textit{tangential flow  Markstein number}, respectively. We note that the term involving $\mc M_c$ in \eqref{burningrate} captures the local effect of flame curvature associated with the propagation of a curved flame relative to the unburnt gas immediately ahead of it; indeed, $1-\mb v \cdot \mb n$ is the non-dimensional flame propagation speed with respect to that gas. The term involving $\mc M_t$ accounts for the effect of tangential flow non-uniformity along the flame front, which induces straining and thus contributes to the flame stretch. Extending these results, in the presence of gravity, formulas~\eqref{burningrate2} and~\eqref{burningrate} are found to generalise to
\begin{align}
     \dot m =
     1- \mc M_c \mathbb K + (\mc M_c - \mc M_t)\nabla_t\cdot\mb v_t    - \mc M_g \nabla_t\cdot\mb g_t\label{burningrate4}  
\end{align}
or, equivalently, 
\begin{align}
    \dot m = 1 + \mc M_c (1-\mb v\cdot \mb n)\nabla\cdot\mb n - \mc M_t \nabla_t\cdot\mb v_t  - \mc M_g \nabla_t\cdot\mb g_t ,\label{burningrate3}
\end{align}
where $\mc M_g$ is a Markstein number characterising the straining induced by gravity along the flame front. Here $\nabla_t\cdot\mb g_t$ is the surface divergence of the tangential gravity component $\mb g_t = \mb g - (\mb g \cdot \mb n) \mb n $, which for usual constant $\mb g$ reduces to $\nabla_t\cdot\mb g_t = -(\mb g\cdot \mb n)\nabla\cdot\mb n$. It is important to point out that the term $\mc M_g \nabla_t\cdot\mb g_t$ is absent in conventional flames modelled by the Navier--Stokes equation. This term, attributable to the use of Darcy's law which is  suitable under strong confinement, will be discussed further below. 

\begin{figure}[h!]
\centering
\includegraphics[width=0.36\textwidth]{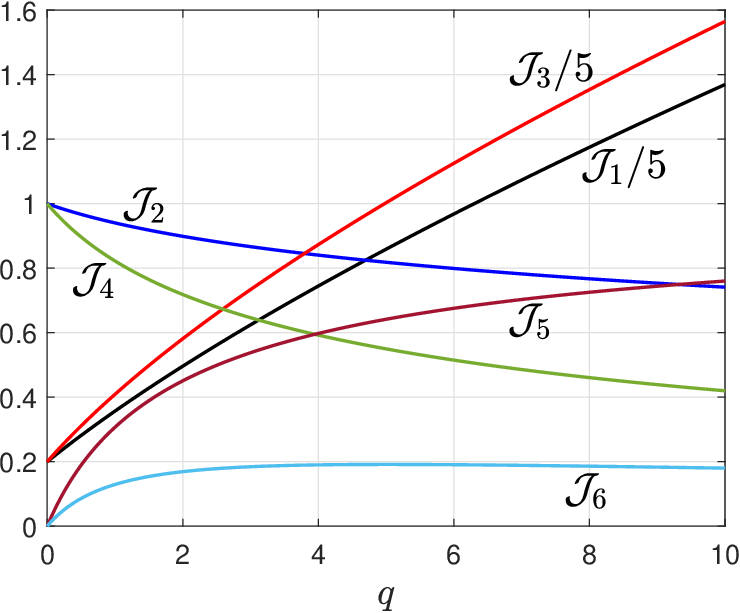} 
\vspace{10 pt}
\caption{The six integrals $\mc J_i$,  defined in \eqref{J1}-\eqref{J6} and appearing in formulas~\eqref{marksteinc}-\eqref{marksteing}, are plotted versus $q$ for the case $\rho=1/(1+q\theta)$ and $\mu=\lambda=(1+q\theta)^{0.7}$.} 
\label{fig:integrals}
\end{figure}

\begin{figure*}[h!]
\centering
\includegraphics[width=0.36\textwidth]{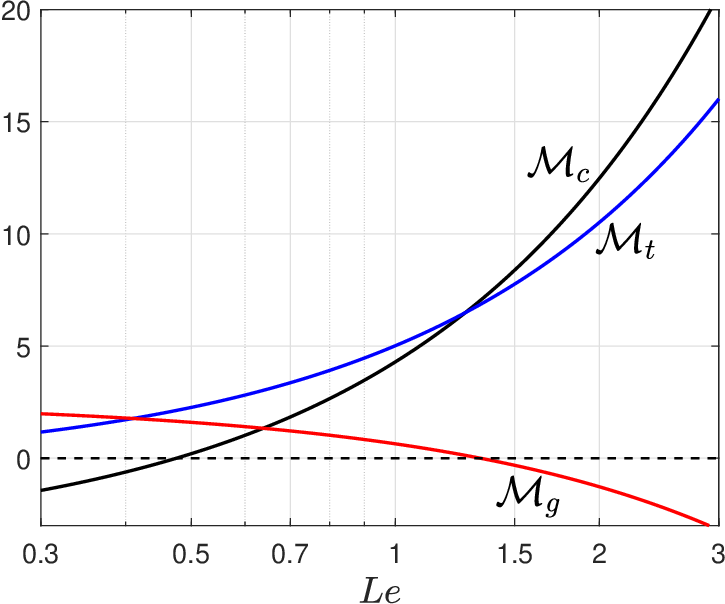}\hspace{1.5cm}
\includegraphics[width=0.36\textwidth]{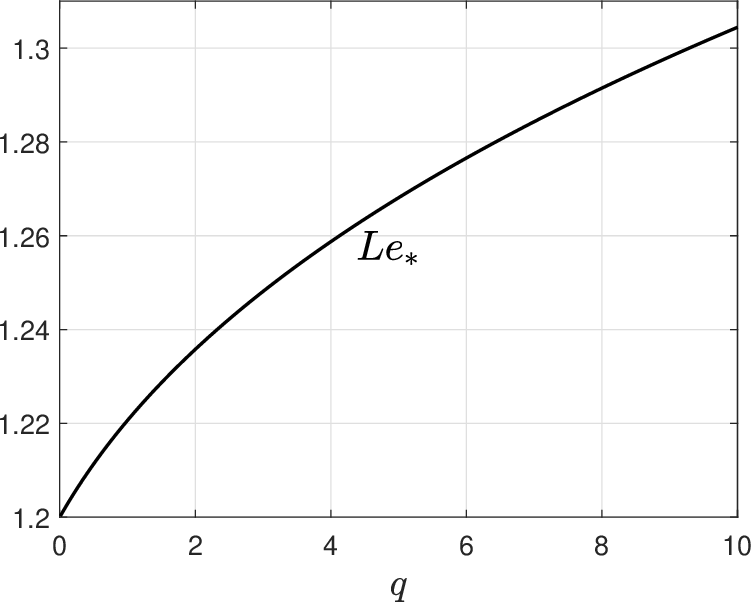} 
\vspace{10 pt}
\caption{Graphical illustration of the three Markstein numbers~\eqref{marksteinc}-\eqref{marksteing} as functions of the Lewis number $Le$ when $\beta=20$ and $q=5$ and for $\rho=1/(1+q\theta)$ and $\mu=\lambda=(1+q\theta)^{0.7}$. The critical Lewis number $Le_*=1+l_*/\beta$ below which $\mc M_t>\mc M_c$, defined in~\eqref{lstar}, is shown on the right as a function of $q$.} 
\label{fig:lstar}
\end{figure*}

We now present the explicit formulas we have derived for $\mathcal{M}_c$, $\mathcal{M}_t$, and $\mathcal{M}_g$. They generalise analogous formulas obtained (with slightly different notation) for the special case of unity Lewis number, $Le = 1$, in \cite{rajamanickam2024hydrodynamic}, and are given by
\begin{align}
    \mathcal{M}_c & = \mathcal{J}_1 + \frac{l}{2}\mathcal{J}_2 , \label{marksteinc} \\
    \mathcal{M}_t &= \mathcal{J}_3 + \frac{l}{2}\mathcal{J}_4 ,\label{marksteint}\\
    \mathcal{M}_g &= \mathcal{J}_5 - \frac{l}{2}\mathcal{J}_6 . \label{marksteing}
\end{align}
Here $l=\beta(Le-1)$ is the reduced Lewis number, $\beta$ is the Zeldovich number, and the six quantities $\mathcal{J}_i$ are the integrals given by 
\begin{align}
    \mc J_1 &= \int_0^1 \frac{\lambda}{\theta}[1-\rho(1-\theta)] d\theta, \label{J1}\\
    \mc J_2 &= -\int_0^1\rho\lambda \ln\theta d\theta, \\ \label{J3}
  \mc J_3 &= \int_0^1 \frac{\lambda}{\theta}\left[1-\frac{\rho}{\mu}(1-\theta)\right] d\theta \,, \\
  \label{J4}
  \mc J_4 &= - \int_0^1\frac{\rho}{\mu}\lambda \ln\theta d\theta \,, \\
  \mc J_5 &= \int_0^{1} \frac{\rho\lambda}{\mu\theta}(1-\rho)(1-\theta) d\theta \,, \\
   \mc J_6 &=  -\int_0^1\frac{\rho\lambda}{\mu}(1-\rho)\ln\theta d\theta. \label{J6}
\end{align}
The integrals $\mc J_i$, which are positive, are plotted in Fig.~\ref{fig:integrals} as functions of $q$. We note that  in the weak heat-release approximation ($q\ll 1$),  
$\mc M_c$ and $\mc M_t$ reduce to  $1+\tfrac{1}{2}l+O(q)$ and $\mc M_g \sim O(q)$, thereby recovering the familiar formula $\dot m = 1 - (1+\tfrac{l}{2})\mathbb K$ applicable in  the thermo-diffusive approximation of constant density and constant transport properties.

\begin{figure*}[h!]
\centering
\includegraphics[width=0.7\textwidth]{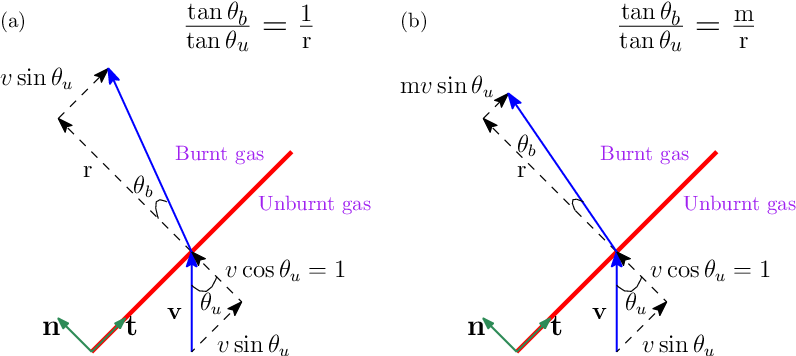} 
\vspace{10 pt}
\caption{Refraction of a streamline across a stationary oblique flame front: (a) continuous tangential velocity (usual case based on Navier-Stokes); (b) discontinuous tangential velocity with $\mathrm{g}_t=0$ (Darcy's law).}
\label{fig:streamline}
\end{figure*}

Formulas~\eqref{burningrate2}-\eqref{marksteing} may be compared with the corresponding formulas for conventional (unconfined or weakly confined) flames, namely the Clavin–Williams formulas \cite{clavin1982effects},
\begin{equation}
\mc M_c = \mc M_{t} = \mc J_1 + \frac{l}{2}\mc J_2, \quad \dot m = 1-\mc M_c \mathbb K. \label{clavinwilliams}
\end{equation}
The most notable difference is the inequality of the two Markstein numbers $\mc M_c$ and $\mc M_t$ under Darcy's law, along with the appearance of a third Markstein number $\mc M_g$, which is absent in \eqref{clavinwilliams} (the latter being based on the Navier–Stokes equation). At the root of this distinctive feature is the fact that Darcy's law permits leading-order discontinuities in the tangential velocity across the flame; see subsection~3.1.  Consequently, the kinematic stretch $\mathbb K$ alone is insufficient to determine the burning rate in this context, except in problems where the tangential straining is zero, i.e.,
\begin{equation}
\nabla_t\cdot \mathbf{v}_t = \nabla_t\cdot \mathbf{g}_t = 0. \label{Darcycondition}
\end{equation}
A further distinction  lies in the dependence of the    Markstein numbers on transport properties. While $\mc M_c$ depends on the density and thermal conductivity profiles $\rho(\theta)$ and $\lambda(\theta)$, it is independent of viscosity variations within the flame. For instance, setting $\mu=1$ while retaining $\rho=\rho(\theta)$ and $\lambda=\lambda(\theta)$ leaves $\mc M_c$ unchanged. By contrast, the tangential-flow Markstein number $\mc M_t$ relies crucially on viscosity variations across the flame, and $\mc M_g$, though to a lesser extent, also exhibits a non-negligible dependence on such variations.

We now comment on the Markstein numbers. Their physical meanings are encapsulated in the formulas~\eqref{burningrate4} and~\eqref{burningrate3}, and their defining expressions in formulas~\eqref{marksteinc}-~\eqref{marksteing}. These formulas are used to  generate the left plot in Fig.~\ref{fig:lstar}.  We note    that $\mc M_c,\mc M_t$ and $\mc M_g$ are all positive when $Le$ is near unity.    Furthermore,   the expressions derived imply   that $\mc M_c$  and $\mc M_t$ are unequal unlike  in  the traditional case, as mentioned above, and  more specifically that 
\begin{equation}
    \mc M_t \gtrless \mc M_c \quad \text{if} \quad l \lessgtr l_*  \equiv 2\frac{\mc J_3-\mc J_1}{\mc J_2-\mc J_4} \,.\label{lstar}
\end{equation}
The reduced Lewis number $l_*$ is used to generate the right plot of $Le_* = 1+ l_*/\beta$ versus $q$ in Fig.~\ref{fig:lstar}. We note that $Le_*$ is always larger than unity, since $l_*\geq 4$ which is a consequence of $l_*$ being an increasing function of $q$ with $l_*\to 4$ as $ q\to 0$.
For $l=l_*$, these two Markstein numbers have equal contributions  leading to the classical relationship  $\dot m = 1 - \mc M_c \mathbb K$ 
   in~\eqref{clavinwilliams}. In contrast, as $l$ is increased above $l_*$,  $\mc M_t$  decreases with respect to  $\mc M_c$, leading to a decrease of the tangential strain contribution relative to that of curvature in the burning-rate formula~\eqref{burningrate3}. 
   
To close this section, we will assess the gravity term's relevance using practical values. For the Hele-Shaw channel, from the definition of $|\mb g|$ given earlier, 
\begin{equation}
    |\mb g|= \frac{\ep Pr}{12 Fr^2}, \quad \ep = \frac{h}{\delta_L} , \label{gravityepsilon}
\end{equation}
with Prandtl number $Pr$ and Froude number $Fr=S_L/\sqrt{gh}$. In the vertical-channel experiments of~\cite{veiga2020thermoacoustic}, the values $Fr^2\sim 10^{-3}$-$1$ are reported  and $\ep \sim 1$-$10^2$. A lower-bound estimate for $|\mb g|$ with $\ep \sim 1$ gives $|\mb g| \sim 10^{-1}$, indicating that gravity effects may not be negligible.

\section{Application to canonical configurations}

The hydrodynamic model is now illustrated for a few canonical configurations, using equations~\eqref{darcyunburnt}–\eqref{kinematic} together with the burning-rate formula~\eqref{burningrate3}.

\subsection{Steady, oblique flame front - Augmented streamline refraction} \label{sec:ob} 

A key difference between Navier–Stokes hydrodynamics and Darcy's hydrodynamics is the allowance of tangential discontinuities in the latter, as can be deduced from~\eqref{eq:FullDarcy}. Indeed, imposing pressure continuity $\llbracket p \rrbracket=0$ in the tangential component of~\eqref{eq:FullDarcy} yields
\begin{equation}
\llbracket \mu \mb v_t \rrbracket = \llbracket \rho \mb g_t\rrbracket \,.
\end{equation}
This implies $\llbracket \mb v_t \rrbracket \neq 0$, arising either from the viscosity jump $\llbracket \mu\rrbracket = 1/\mm - 1$ or from the gravitational jump $\llbracket \rho \mb g_t\rrbracket= (1/\rr-1)\mb g_t$ across the flame.

The tangential discontinuity in $\mb v_t$ has an important consequence for oblique  flame fronts. Typically, across such fronts, streamlines deflect towards the normal $\mb n$ upon crossing due to the increase in $v_n=\mb v\cdot \mb n$,  the refraction being characterised by
\begin{equation} \label{refra1}
\frac{\tan \theta_b}{\tan \theta_u} = \frac{1}{\rr},
\end{equation}
with the notation corresponding to the planar flame sketched in  Fig.~\ref{fig:streamline}(a). However, under Darcy's law, in addition to this deflection associated with the increase in  the normal component $v_n$, there is also a decrease in the tangential component  $v_t =\mb v \cdot \mb t$, which leads to augmented refraction.  This is illustrated  in Fig.~\ref{fig:streamline}(b) in the particular case $\mb g = \mb 0$, for which the refraction is described  by
\begin{equation} \label{refra2}
\frac{\tan \theta_b}{\tan \theta_u} = \frac{\mm}{\rr}.
\end{equation}
Consequently, since $\mm \approx 1/3$, formulas~\eqref{refra1} and~\eqref{refra2} imply that streamlines refract about three times more under Darcy's law than in the classical case.

With gravity being accounted for, the tangential burnt gas velocity in Fig.~\ref{fig:streamline}(b) becomes $\mm v\sin\theta_u - \mathrm{g}_t(\rr-1)\mm/\rr$, meaning that gravity can either aid or oppose refraction depending on the sign of $\mathrm{g}_t = \mb g \cdot \mb t$.  Formula~\eqref{refra2} thereby generalises to
\begin{equation}
\frac{\tan \theta_b}{\tan \theta_u} = \frac{\mm}{\rr}\left[1 - \frac{(\rr-1) \mathrm{g}_t}{\rr v \sin \theta_u} \right].
\label{eq:gravity_refraction}
\end{equation}
 When $ \mathrm{g}_t<0$ in Fig.~\ref{fig:streamline}, corresponding to a stable configuration with unburnt gas below burnt, gravity reduces streamline refraction. Conversely, when $\mathrm{g}_t>0$, representing an unstable configuration with burnt below unburnt (associated with Rayleigh–Taylor instability), the deflection is amplified.

To close this subsection, we note that augmented streamline refraction under Darcy’s law is expected to have implications for hydrodynamic flame instability, as qualitative explanations of this instability typically invoke such refraction effects~\cite[pp.58-59]{clavin2016combustion}. These stability aspects are addressed in Subsection~3.4.

\subsection{Radially symmetric flames}

\begin{figure}[h!]
\centering
\includegraphics[width=0.15\textwidth]{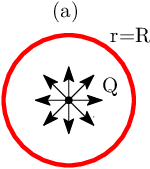} 
\includegraphics[width=0.17\textwidth]{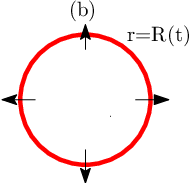} 
\vspace{10 pt}
\caption{Radially symmetric flame configurations: (a) stationary flame in a point-source flow of reactive gas of strength $Q$ and (b) freely propagating flame.}
\label{fig:radial}
\end{figure}

Consider now the illustrative case of radially symmetric flames, such as those shown in Fig.~\ref{fig:radial}. Whether steady or evolving in time, inwardly or outwardly propagating, the velocity field on each side of the flame is purely radial and, by incompressibility, takes the form
\begin{equation}
  v_r \propto \frac{1}{r^{d-1}}, \qquad d=2,3 ,
\end{equation}
where $d$ denotes the spatial dimension. Since the tangential velocity vanishes identically ($\mb v_t=0$), the burning rate reduces to $\dot m = 1 -\mc M_c\mathbb K$. This configuration yields the same burning rate under Darcy and Navier–Stokes hydrodynamics, though the radial pressure distribution differs between the two.

\begin{figure}[h!]
\centering
\includegraphics[width=0.16\textwidth]{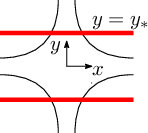} 
\vspace{10 pt}
\caption{Twin premixed flame in a planar counterflow.}
\label{fig:stag}
\end{figure}

\subsection{Twin premixed counterflow flame}

The counterflow configuration is particularly instructive for illustrating our Darcy-law-based model, as it involves tangential straining of the flame, thereby activating the term $\mc M_t \nabla_t \cdot \mb v_t$ in the burning-rate formula~\eqref{burningrate3}. Consider the twin premixed flame configuration in a two-dimensional, planar counterflow, illustrated in Fig.~\ref{fig:stag}, with $\mb g=0$ for simplicity. The unburnt gas occupies the region $|y|>|y_*|$, and the burnt gas the region $|y|<|y_*|$.

By symmetry, it suffices to consider the half-domain $y>0$. The hydrodynamic model~\eqref{darcyunburnt}–\eqref{kinematic} yields the solution
\begin{align} 
    y_*<y<\infty: &\quad v_x = A x, \quad v_y = -A (y-d), \nonumber\\ \nonumber
    &\quad p= \frac{A}{2}(y^2-x^2) - Ad y, \\ 
    0<y<y_*: &\quad v_x = \mm A x, \quad v_y = -\mm A y, \nonumber\\ &\quad p= \frac{A}{2}(y^2-x^2) - Ady_*, \nonumber
\end{align}
where $A$ is the non-dimensional strain rate in the unburnt gas and $d$ is a constant, with $y=d$ representing the apparent stagnation plane location as seen from the unburnt gas.  The parameters $d$, $y_*$, and the burning rate $\dot m$ are given by
\begin{equation*} 
    d = y_*-\frac{\dot m}{A}, \quad y_* = \frac{\rr \dot m}{\mm A}, \quad \dot m = 1-\mc M_t A. 
    \end{equation*} 
This solution invites several interesting observations:

(a)   The jump in normal velocity at the flame is $\llbracket v_y \rrbracket = -(\rr-1)\dot m$, as expected from mass conservation regardless of whether the momentum equation is Navier–Stokes or Darcy. On the other hand, the tangential velocity is also discontinuous here, with $\llbracket v_x \rrbracket = -(1-\mm)A x$ at the flame front, a distinctive feature of Darcy's law.

(b) Unlike conventional counterflow flames, where the strain-rate jump $\llbracket \partial v_y/\partial y \rrbracket = (\sqrt{\rr}-1)A$ is set by the density ratio $\rr$~\cite{weiss2017aerodynamics}, here it is $\llbracket \partial v_y/\partial y \rrbracket = (\mm-1)A$, governed by the viscosity/permeability ratio $\mm$. In this sense, the Hele-Shaw counterflow is fundamentally different from its classical counterpart.

(c) The pressure field is hyperbolic, quadratic in $x$ and $y$ with opposite signs, everywhere in the domain. This stands in stark contrast to conventional counterflow flames, where the pressure is circular (quadratic with the same sign) in the unburnt gas while being more complicated in the burnt region due to rotational flow. In Darcy's law, by comparison, vorticity is zero in both the unburnt and burnt gas away from the flame~\cite{daou2025hydrodynamic}.

(d) Within the hydrodynamic model, premixed flame extinction can be defined by the condition $\dot m = 0$, or equivalently $y_* = 0$, yielding an extinction strain rate $A_{\mathrm{ext}} = 1/\mc M_t$. Since $\mc M_t$, given in~\eqref{marksteint} together with~\eqref{J3} and~\eqref{J4}, includes additional contributions from viscosity variations ($\mu \neq 1$), this indicates that strained premixed flames are more prone to extinction under Darcy's law whenever $l < l_*$ (see~\eqref{lstar}) than in the conventional counterflow configuration. We may therefore anticipate, for example, that a planar strained hydrogen flame may not be able to sustain strain rates as high as those  in conventional counterflows before cellular breakup occurs~\cite{zhou2023propagation,carpio2020near}.

\subsection{Evolution of unstable planar flames in the presence of a uniform flow}\label{sec:planarstability}

Consider a two-dimensional planar flame propagating in the negative $y$-direction in the presence of a uniform oncoming flow of non-dimensional magnitude $\mc V$ (measured with $S_L$); the flow opposes the flame when $\mc V>0$ and aids it when $\mc V<0$. The non-dimensional gravity vector is taken as $\mb g =   \mathrm{g} \mb e_y$ where $\mathrm{g} < 0$ when $\mb g$ points toward the unburnt gas and $\mathrm{g} > 0$ otherwise. The undisturbed  flame front is planar and is located at $G = y+(1-\mc V)t=0$; the  flame is thus stationary when $\mc V=1$ (that is when the dimensional incoming flow speed is $S_L$) which is the   case studied by Joulin \& Sivashinsky~\cite{joulin1994influence}. The flow field, as in~\cite{daou2025hydrodynamic}, is given by
\begin{align}
    G<0: &\,\, \mb v=\mc V\mb e_y, \,\, p = -\mc V y + \mathrm{g}y, \label{base1} \\ 
    G>0: &\,\, \mb v = (\mc V + \rr -1) \mb e_y, \,\, p = - \frac{(\mc V+\rr-1)}{\mm} y + \frac{ \mathrm{g}}{\rr}y. \label{base2}
\end{align}

A  stability analysis of this planar flame is carried out in our recent work~\cite{daou2025hydrodynamic}. The main outcome of the analysis is the derivation of the dispersion relation  
\begin{equation}
s = \frac{a k-bk^2}{1+c k},    \label{darcydisp}
\end{equation}
determining the growth rate $s$ of perturbations of wavenumber $k$ (assumed positive).    Here
\begin{align}   
 a &= \frac{\rr-1}{1+\mm} + \frac{1-\mm}{1+\mm}\mc V + \frac{\rr-1}{1+\mm}\frac{\mm}{\rr}\mathrm{g}, \label{formula1} \\
     b &=  \left[\frac{2\rr}{1+\mm} + \frac{1-\mm}{1+\mm}(\mc V-1) + \frac{\rr-1}{1+\mm}\frac{\mm}{\rr}\mathrm{g}\right]\mc M_t  +  \frac{\rr+\mm}{1+\mm} \mathrm{g}\mc M_g   ,\label{formula2} \\
       c&=\frac{\rr-1}{1+\mm}\mc M_t. \label{formula3}
\end{align}
Below is a summary of the main implications of the dispersion relation~\eqref{darcydisp}.

(a) First, the positivity of the parameter $a>0$ in~\eqref{darcydisp} implies a long-wave instability. The three terms defining $a$ in~\eqref{formula1} capture, respectively, the influence of the Darrieus--Landau (DL), Saffman--Taylor (ST), and Rayleigh--Taylor instabilities. Each term is individually destabilising when positive and stabilising otherwise.

(b) The positivity of the parameter $b$ corresponds to physically admissible Markstein-type stabilisation, whereby curvature and tangential strain act to smooth short-wavelength perturbations. The requirement $b>0$ is thus necessary; should it be violated, the model would require revision. This condition imposes a constraint on the admissible values of the parameters involved, such as $\mc V$, $|\mb g|$, and the Markstein numbers.

(c) The maximum growth rate occurs at the wavenumber $k_m=(\sqrt{1+ac/b}-1)/c$ with growth rate value given by $s_m=bk_m^2$. The range of unstable wavenumbers is $0<k<a/b$.

(d) A key observation is that the dispersion relation no longer involves the curvature Markstein number $\mc M_c$ as defined in formula~\eqref{burningrate3}. This is because the factor $(1-\mb v \cdot \mb n) \nabla \cdot \mb n$ vanishes then to linear order in the stability analysis, leaving only $\mc M_t$ and $\mc M_g$ contributions. Another aspect worth emphasising is that viscosity variations ($\mu \neq 1$) are ultimately responsible for the dependence of the dispersion relation on the uniform flow magnitude $\mc V$; indeed, when $\mu = 1$, we have $\mc M_c = \mc M_t$ and $\mc V$ disappears from equations~\eqref{darcydisp} to~\eqref{formula3}. Consequently, when $\mu = 1$, the linear stability of the flame becomes insensitive to the presence of a uniform flow, recovering the usual behaviour common to unconfined flames.

\vspace{0.1cm}

We now highlight the fundamental difference between the stability of freely propagating flames ($\mc V=0$) and that of a flame opposed by an imposed flow field, as in the case $\mc V=1$ considered in~\cite{joulin1994influence}.
For the freely-propagating case ($\mc V=0$), we have
\begin{align}
    a &= \frac{\rr-1}{1+\mm} + \frac{\rr-1}{1+\mm}\frac{\mm}{\rr}\mathrm{g},\qquad c=\frac{\rr-1}{1+\mm}\mc M_t, \nonumber  \\
     b &= \left(\frac{2\rr}{1+\mm} - \frac{1-\mm}{1+\mm} + \frac{\rr-1}{1+\mm}\frac{\mm}{\rr}\mathrm{g}\right)\mc M_t  +  \frac{\rr+\mm}{1+\mm} \mathrm{g}\mc M_g. \nonumber 
\end{align}
On the other hand, for the case considered by Joulin and Sivashinsky~\cite{joulin1994influence}, $\mc V=1$, we have
\begin{align}
    a &= \frac{\rr-\mm}{1+\mm}  + \frac{\rr-1}{1+\mm}\frac{\mm}{\rr}\mathrm{g},\qquad c=\frac{\rr-1}{1+\mm}\mc M_t, \nonumber \\
     b &= \left(\frac{2\rr}{1+\mm}  + \frac{\rr-1}{1+\mm}\frac{\mm}{\rr}\mathrm{g}\right)\mc M_t   +  \frac{\rr+\mm}{1+\mm} \mathrm{g}\mc M_g . \nonumber 
\end{align}

\subsection{Weakly nonlinear regime: the Michelson--Sivashinsky equation for Darcy flows}
\addvspace{10pt}

Near the instability threshold ($a \gtrsim 0$), the unstable wavenumbers scale as $k = O(a)$ when $a \to 0^+$, and the growth rate becomes $s \approx a |k| - b k^2 = O(a^2)$, with $b=(\mc M_t+\mathrm{g}\mc M_g)(\rr+\mm)/(1+\mm)$ remaining $O(1)$. This scaling mirrors that of the classical Darrieus--Landau instability, but the coefficients now incorporate Darcy-specific contributions. Following the approach of Sivashinsky~\cite{ashinsky1988nonlinear,michelson1977nonlinear}, one obtains a Michelson--Sivashinsky (MS) equation for the flame shape deviation $f(x,t)$ from its unperturbed planar value $y=(\mc V-1)t$, namely
\begin{equation}
f_t + \tfrac{1}{2}f_x^2 - b f_{xx} - a \mathbb{I} (f)=0, \label{MS}
\end{equation}
where $\mathbb I\{e^{ikx}\}=|k|e^{ikx}$ denotes the non-local integral operator. This equation captures the interplay between nonlinear steepening, diffusive smoothing, and hydrodynamic destabilisation, with all coefficients now modified by the Darcy context.

\section{Flame stability based on Euler--Darcy's law}

Up to now, our analysis has been restricted to strongly confined systems where Darcy's law applies, such as low-permeability porous media or narrow Hele-Shaw cells. We now turn to the role of confinement, characterised, for instance, by the channel width, on the evolution of planar flame fronts. Following Joulin and Sivashinsky~\cite{joulin1994influence}, we adopt the Euler–Darcy framework, noting that a complete hydrodynamic thin-flame model based on this description has yet to be developed. Within this context, we examine the linear stability of planar flames and their nonlinear evolution, focusing on nearly planar fronts.

The dispersion relation derived in~\cite{daou2025hydrodynamic}, which connects the wavenumber $k$ and growth rate $s$ in the presence of an imposed uniform flow, is given by
\begin{equation}
    c_2(k) s^2 + c_1(k)s + c_0(k)=0,  \label{disp}
\end{equation}
where
\begin{align*}
    c_2&= \frac{\rr+1}{2} + \frac{\rr-1}{\rr}\mc M_t k,\\
    c_1 &= 2k +  2\rr\mc M_tk^2 + \varphi(1+ck),\\
    c_0 &= - a \varphi   k + \big(   b \varphi+  1- \rr    \big)  k^ 2  + (3\rr-1)\mc M_t k^3, 
\end{align*}
with
\begin{align}
\varphi = \frac{12\Pra(1+\mm)}{\ep^2 \mm}, \quad 
      \quad \ep = \frac{\sqrt{12\kappa_u}}{\delta_L}. 
\end{align}
For Hele-Shaw cells, we have $\ep=h/\delta_L$, as defined in~\eqref{gravityepsilon}. Under strong confinement or low permeability ($\ep \to 0$, hence $\varphi \to \infty$), the dispersion relation~\eqref{disp} reduces to the Darcy dispersion relation~\eqref{darcydisp}. In the opposite limit of weak confinement ($\ep \to \infty$, hence $\varphi \to 0$) with $\mc V = 0$, it recovers the classical dispersion relation (see e.g.~\cite{creta2011strain}) governed by Euler's equation. Thus, expression~\eqref{disp} provides a unified description that bridges the two extremes: unconfined flames on one hand and strongly confined, friction-dominated flames on the other.

The onset criterion for instability of the planar flame is now given as a function of $\ep$ (or $\varphi$). Following~\cite{daou2025hydrodynamic}, we assume $\mc M_t>0$ (so that $c_1,c_2>0$). The flame is unstable when $a > a_c$, with
\begin{align}
    a_c &=
    \begin{cases}
       0 & \text{if  } \varphi > \varphi_c, \\ 
       \displaystyle \frac{-b(\rr-1)}{4(3 \rr-1)\mc M_t} \, \frac{\varphi_c}{\varphi}
       \left(1 - \frac{\varphi}{\varphi_c}\right)^{\!2} & \text{if  } \varphi < \varphi_c ,
    \end{cases} 
\end{align}
and $\varphi_c = (\rr-1)/b$. The  critical wavenumber  is
\begin{align}
    k_c &=
    \begin{cases}
       0 & \text{if  } \varphi > \varphi_c, \\ 
       \displaystyle \frac{\rr-1}{2(3 \rr-1)\mc M_t} \left(1 - \frac{\varphi}{\varphi_c}\right) & \text{if  } \varphi < \varphi_c.
    \end{cases} 
\end{align}
The first criterion, $a > a_c=0$, applies under strong confinement (large $\varphi$) and is predicted by Darcy's law; it remains valid up to the critical value $\varphi_c$. For $\varphi < \varphi_c$, the wavenumber at onset is non-zero, with $k_c \to (\rr-1)/2(3 \rr-1)\mc M_t$ as $\varphi/\varphi_c \to 0$, corresponding to unconfined flames.

In the weakly nonlinear regime near threshold ($a \to a_c^+$), the growth rate takes the form 
\begin{equation}
s = \begin{cases}
\displaystyle a |k| - b\left(1-\frac{\varphi_c}{\varphi}\right) k^2 & \text{if } \varphi > \varphi_c, \\
\displaystyle \alpha (a-a_c) - \gamma (k-k_c)^2 & \text{if } \varphi < \varphi_c,
\end{cases} \label{dispapprox}
\end{equation}
where $\alpha = \varphi k_c / c_1(k_c)$ and $\gamma = (3 \rr-1)\mc M_t k_c / c_1(k_c)$ are positive numbers.
For $\varphi > \varphi_c$, the flame dynamics is governed by a Michelson–Sivashinsky(MS)-type equation~\eqref{MS}, with $b$ replaced by $b(1-\varphi_c/\varphi)$, namely
\begin{equation}
f_t + \tfrac{1}{2}f_x^2 - b\left(1-\frac{\varphi_c}{\varphi}\right) f_{xx} - a \mathbb{I}(f)=0.
\end{equation}
This equation ceases to be valid as $\varphi \to \varphi_c^+$, since the coefficient   of $f_{xx}$ vanishes. In this limit, a cubic term $k^3$ must be retained  in~\eqref{dispapprox}. Although the MS equation is derived for $a\ll 1$, it is expected to qualitatively reproduce the flame dynamics such as the coalescence of poles. For $\varphi < \varphi_c$, the non-zero critical wavenumber $k_c$ points to Ginzburg–Landau-type dynamics~\cite[pp.217-218]{cross2009pattern} and the flame front exhibits a dominant Fourier mode at $k = k_c$ with slowly modulated amplitude, governed by the   Ginzburg–Landau (GL) equation.
Explicitly, setting
\begin{equation}
f = \frac{3}{2k_c}\sqrt{\frac{(a-a_c)\alpha}{\gamma}}  \,  \text{Re} \left\{A(X,T)e^{ik_c x}\right\}
\end{equation}
leads to the real Ginzburg–Landau equation
\begin{equation}
A_T = A_{XX} + A - |A|^2 A \label{GLequation}
\end{equation}
involving  the slow variables  $X=x\sqrt{(a-a_c)\alpha/\gamma}$ and $T=t(a-a_c)\alpha$.  The GL description remains valid as long as $\varphi$ is finite, corresponding to moderately confined channels. In the unconfined limit $\varphi \to 0$, the second expression in~\eqref{dispapprox} reduces to
\begin{equation}  \label{dispLand}
s = \frac{\rr-1}{2}|k| - \mc M_t k^2 ,
\end{equation}
since $\alpha \to 0$, $\gamma \to \mc M_t$, and $k_c \to (\rr-1)/4\mc M_t$ with the instability onset now identified by $\rr = 1^+$. The dispersion relation~\eqref{dispLand} then leads to the classical Michelson–Sivashinsky equation
\begin{equation}
f_t + \tfrac{1}{2}f_x^2 - \mc M_t f_{xx} - \frac{\rr-1}{2} \mathbb{I}(f) = 0.
\end{equation}
It is instructive to summarise the stability results for $\mc V=0$ (freely propagating flame with no imposed flow) and $g=0$. In the unconfined limit, where Euler's equation is applicable, the  DL growth rate scales as $s \sim |k|(\rr-1)/2$ for long waves. In the strongly confined Darcy limit, on the other hand, one finds $s \sim |k|(\rr-1)/(1+\mm)$. Since $1+\mm < 2$, confinement is seen to amplify the DL instability.  This enhancement can be traced back to the amplified streamline refraction discussed in subsection~3.1.

\section{Conclusions}

This study analyses confined premixed flames governed by Darcy's law, such as in porous media and Hele-Shaw cells. A major finding is the existence of two distinct Markstein numbers, for curvature and tangential strain, whose inequality is a distinctive feature of Darcy's law absent in conventional flames. This  feature vanishes in purely radial flows, where tangential straining is absent, but fundamentally modifies the flame response in configurations with tangential flow, such as the planar counterflow. Notably, in such flows, the strain rate experienced by the flame is governed by the viscosity ratio $\mm$ rather than the density ratio $\rr$, a dramatic departure from conventional counterflow flames. A third Markstein number $\mc M_g$, associated with gravity-induced straining, also emerges uniquely under Darcy's law. This Markstein number may become significant in vertical channels.

Stability analysis reveals that confined planar flames experience combined Darrieus--Landau, Saffman--Taylor, and Rayleigh--Taylor instabilities.   Weakly nonlinear analysis shows that strongly confined flames follow a Michelson--Sivashinsky equation with confinement-modified coefficients, while moderately confined flames exhibit Ginzburg--Landau dynamics. Strong confinement amplifies the Darrieus--Landau instability, an effect traced to augmented streamline refraction caused by the tangential velocity discontinuity permitted by Darcy's law. Future work could incorporate heat losses~\cite{keller1994transient,matalon2009multi}, transient-pressure variations~\cite{keller1994transient,bechtold2024hydrodynamic}, and acoustics~\cite{searby1991parametric,bychkov1999analytical,law2013effect,veiga2020thermoacoustic,veiga2019experimental} within Darcy’s law.

\section*{Acknowledgments} 

This work was supported by UK EPSRC through Grant No.~APP39756.

\bibliographystyle{elsarticle-num}
\bibliography{elsarticle-template}

\end{document}